%% file: main.tex
\documentclass{pasj00}

\begin{document}
\SetRunningHead{Deguchi et al.}{SiO Maser Survey in the Inner Galactic Disk}
\Received{2004/05/06}
\Accepted{2004/07/09}

\title{SiO Maser Survey of IRAS Sources in the Inner Galactic Disk}

\author{Shuji \textsc{Deguchi}$^{1}$,  
        Takahiro \textsc{Fujii}$^{2,3}$, Ian S. \textsc{Glass}$^{4}$, 
Hiroshi \textsc{Imai}$^{2,3,5}$, Yoshifusa \textsc{Ita}$^{6}$,}
\author{Hideyuki \textsc{Izumiura}$^{7}$, Osamu \textsc{Kameya}$^{2, 8}$, 
        Atsushi \textsc{Miyazaki}$^{1}$, Yoshikazu \textsc{Nakada}$^{6}$
}

\and
\author{Jun-ichi {\sc Nakashima}$^{9, 10}$}
\affil{$^{1}$ Nobeyama Radio Observatory, National Astronomical Observatory,\\
              Minamimaki, Minamisaku, Nagano 384-1305}
\affil{$^{2}$ VERA Project Office, National Astronomical Observatory, 
              2-21-1 Osawa, Mitaka, Tokyo 181-8588}
\affil{$^{3}$ Faculty of Science, Kagoshima University, 
              1-21-35 Korimoto, Kagoshima 890-0065}
\affil{$^{4}$ South African Astronomical Observatory,
              PO Box 9, Observatory 7935, South Africa}
\affil{$^{5}$ Joint Institute for VLBI in Europe, 
              Postbus 2,  7990 AA Dwingeloo, The Netherlands}
\affil{$^{6}$ Institute of Astronomy, School of Science, The University of Tokyo,
              2-21-1 Osawa, Mitaka, Tokyo 181-0015}
\affil{$^{7}$ Okayama Astrophysical Observatory, National Astronomical 
Observatory, \\ Kamogata, Asakuchi, Okayama 719-0232}
\affil{$^{8}$ Mizusawa Astrogeodynamics Observatory, National 
Astronomical Observatory, Mizusawa, Iwate 023-0861}
\affil{$^{9}$ Department of Astronomical Science, The Graduate University 
for Advanced Studies,\\
Minamimaki, Minamisaku, Nagano 384-1305}   
\affil{$^{10}$ Department of Astronomy, University of Illinois at Urbana-Champaign\\
1002 W. Green St. , Urbana, IL 61801, U.S.A.}
\author{\\(PASJ  --- 2004 Aug. 25 issue in press)}


\maketitle

\begin{abstract}

We have surveyed 401 color selected  IRAS sources in the Galactic disk in
the SiO $J=1$--0 $v=$ 1 and 2 maser lines at 43 GHz, resulting in 254 (239
new) detections. The observed sources lie mostly in a strip of the inner
Galactic disk with boundaries $-10^{\circ}<l<40^{\circ}$ and $|b|<3^{\circ}$.
This survey provides radial velocities of inner-disk stars for which optical
measurements cannot be made due to interstellar extinction.  The SiO
$l$--$v$ diagram in the area $-10^{\circ}<l<40^{\circ}$ exhibits fewer
objects coincident with the molecular ring feature than the OH 1612 MHz
source $l$--$v$ map does, indicating a slight difference of stellar type
between SiO and OH emitting stars.  After identifying all of the SiO
detected sources in the 2MASS near-infrared catalog, we computed their
luminosity distances based on the infrared fluxes. We then mapped these
objects onto the first quadrant of the Galactic plane. Combining the distances
with the SiO radial-velocities, we obtained a pattern speed for SiO maser
sources, $\Omega _{\rm P}=21 \; (\pm 13)$ km s$^{-1}$ kpc$^{-1}$, between
the distances 1 and 5.5 kpc, without the use of any dynamical models. The
increase of the pattern speed toward the Galactic center (up to 60 km
s$^{-1}$ kpc$^{-1}$ between the distances, 5.5 and 7 kpc) suggests the
presence of two pattern speeds in the Galaxy.
\end{abstract}

\section{Introduction}

Observations of OH 1612 MHz and SiO maser sources (\cite{tel91};
\cite{jia96}; \cite{sev97}) provide basic radial velocity data  for
stars throughout the Galaxy with an accuracy of a few km s$^{-1}$.  The
accumulation of radial-velocity data is useful for studying the rotation
of the inner and outer parts of the Galaxy (\cite{nak00};
\cite{nak03}), and complements the
radial-velocity data based on the HI and CO gas.  At visible wavelengths,
these data are difficult to obtain because of interstellar extinction in the
Galactic disk.  In particular, the radial velocity data can be used to
obtain the pattern speed of the Galaxy without appealing to dynamical models
(\cite{deb02}).

The SiO masers arise mostly in the circumstellar envelopes of mass-losing
evolved stars in the Asymptotic Giant Branch (AGB) phase (or occasionally
the post-AGB phase), which are intrinsically bright in the near- and
mid-infrared regions.  Though many surveys have been made during the last
decade in SiO and OH, a considerable number of candidate stars for masers in
the inner part of the Galactic disk have not previously been observed.  The
present paper aims to improve the radial velocity data in the sky areas
where the maser surveys have been relatively poor.


Though a large number of OH 1612 MHz objects have been detected in the area
$|l|<45^{\circ}$ and $|b|<3^{\circ}$ (\cite{sev01}), SiO maser
sources have not been the subject of deep searches [except for \citet{izu99},
and \citet{deg00b} who looked at objects having a relatively narrow infrared
color range].  A comparison of variables detected in SiO with those detected
in OH 1612 MHz at the Galactic center (\cite{ima02}; \cite{deg04}) revealed
that OH 1612 MHz objects tend to be of longer period than those detected in
SiO; in the period range between 200--500 days, the detection rate of SiO is
triple that of OH. This fact indicates that the SiO maser samples are weighted
towards stars with smaller mass than the OH maser samples, based on
the observed increase of mass with period for Miras (e.g.,
\cite{fea96}).  In addition, the OH 1612 MHz sample always involves some
contamination by young objects associated with star forming regions (SFR),
and therefore molecular clouds, even though they have doubly peaked line
profiles (\cite{cas99}).  These might bias the radial-velocity statistics to
some degree because Galactic gas dynamics are considerably different from
stellar dynamics (see, for example, \cite{bin91}).  In this sense, the SiO
maser radial-velocity sample can rectify such a bend for studying Galactic dynamics. 
Furthermore, in previous surveys (for example, \cite{nak03}), the
overlap of objects with SiO and OH 1612 MHz was only 1/3 of the detected
objects. Therefore, it is useful to increase the number of SiO detections to
enlarge the sample.

Apart from the fact that SiO maser observations of color-selected IRAS
sources provide basic data on radial velocities, they show with certainty
that these objects are O-rich mass-losing evolved stars. Though
classifications of IRAS point sources have been made using IRAS low
resolution spectra and IRAS colors (\cite{kwo97}), the natures of some of
the faint objects with red IRAS colors remain uncertain, i.e., as to
whether they are evolved (AGB/post AGB) stars or young stellar objects. IRAS
colors or even IRAS low resolution spectra (LRS) can often be insufficient
for this purpose (for example, LRS class=$1n$ or $3n$ objects; \cite{oln86}).

In this paper, we present the survey data with minimum interpretation.  We
have now accumulated a large amount of SiO data (403 objects during
the past 8 years of observations); some objects have turned out to be quite
interesting. About 85 percent of the objects in the present paper are in
the area $-10^{\circ}<l<40^{\circ}$ and $|b|<3^{\circ} $ because the surveys
were made as a backup program during two other long-term projects looking
towards the Galactic center.

\section{Observations}

Simultaneous observations in the SiO $J=1$--0, $v=1$ and 2 transitions at
42.122 and 42.821 GHz, respectively, were made with the 45-m radio telescope
at Nobeyama during the period 1997 January--2003 May.  
We used a cooled SIS mixer receiver (S40) for the 43 GHz
observations and accousto-optical spectrometer arrays, AOS-H and AOS-W,
having bandwidths of 40 and 250 MHz. The effective velocity resolution of
the AOS-H spectrometer is 0.3 km s$^{-1}$. They cover the velocity range of
$\pm 350 $ km s$^{-1}$, for both the SiO $J=1$--0 $v=1$ and 2 transitions,
simultaneously. The overall system temperature was between 200 and 300 K,
depending on weather conditions. The half-power telescope beam width (HPBW)
is about 40$''$. The antenna temperature ($T_{a}$) given in the present
paper is that corrected for the atmospheric and telescope ohmic loss but not
for the beam or aperture efficiency. The conversion factor of the antenna
temperature to the flux density is approximately 2.9 Jy K$^{-1}$. We
employed the position switching mode. Further details of SiO maser
observations using the NRO 45-m telescope have been described elsewhere
(\cite{deg00a}), and are not repeated here.

The sources observed in this paper were mainly selected according to the
following criteria:

\begin{enumerate}

\item  IRAS sources in the narrow color range, $0.0 <C_{12}\;
 (=log(F_{25}/F_{12}) <0.1$, in the area, $25^{\circ}<l<40^{\circ}$ and
 $|b|<3^{\circ}$, where $F_{12}$ and $F_{25}$ are the IRAS flux densities 
 at 12 and 25 $\mu$m. The surveys of these sources were made during the
 1997--1999 long-term observations of IRAS sources in the Galactic disk
 (\cite{izu99}).

\item The IRAS sources in the color ranges, $-0.1<C_{12}<0.0$ and
 $0.1<C_{12}<0.2$, in the area, $-10^{\circ}<l<40^{\circ}$ and
 $|b|<3^{\circ}$. These sources were observed in the 2001--2003
 winter--spring seasons during the long-term SiO maser survey of
 large-amplitude variables in the Galactic center (\cite{ima02}).  The
 observed positions were the MSX$\_$5CG source positions (\cite{ega99}) 
 within 20$''$ of IRAS sources, but were selected according to their IRAS colors
for consistency with the previous surveys.

\end{enumerate}

The criteria that were used after 2000 effectively exclude some young
stellar objects in dust clouds; they are very often clustered and the MSX
images frequently resolve those for which the IRAS catalog gives only a
single strong source.  Furthermore, the later-half survey after 2000 was
limited to relatively bright objects ($F_{12}\gtrsim5$ Jy).

\begin{figure}
\vspace{-1.cm}
  \begin{center}
    \FigureFile(100mm,40mm){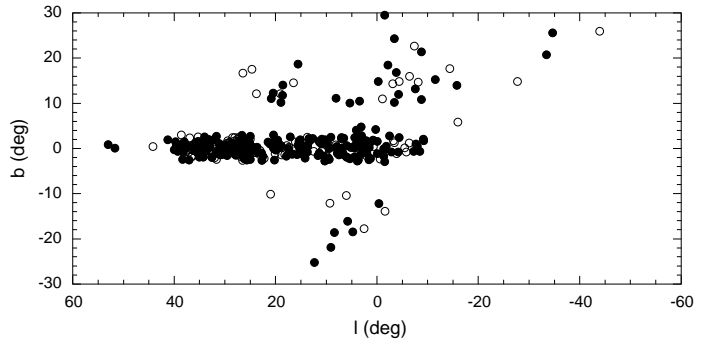}
  \end{center}
  \caption{Source distribution in Galactic coordinates. Filled and unfilled 
circles indicate SiO detections and nondetections.
}\label{fig: l-b map}
\end{figure}

In addition, from time to time, we observed objects at higher galactic
latitudes ($|b|>10^{\circ}$) where the surveys were formerly incomplete. 
These sources were mostly observed as backups during poor weather
conditions, such as when strong wind or thick cloud hindered sensitive
observations. Several low declination sources ($\delta \lesssim
-30^{\circ}$) were also observed during conditions when a part of
the antenna surface was covered with ice (but the weather was mostly
clear). On such occasions, we could observe bright objects at very low
elevation angles, which were normally not considered suitable.


In total, we observed 401 IRAS sources and detected 254 in the SiO $J=1$--0
$v=1$ or 2 transitions in the sky area  $-60^{\circ}<l<60^{\circ}$. 
The results are summarized in tables 1 and 2 for 254 detections and 147
nondetections, respectively.  Previous OH/SiO observations are noted in the
last two columns of table 1 as "y" (detection) and "n" (nondetection), with
references.  Among 254 detected sources, 239 are new SiO detections, and 183
were not detected before in either OH 1612 MHz or SiO maser lines. The
distribution of the observed objects in Galactic coordinates is shown in
figure 1. As seen in this figure, about 85\% of the objects (342 sources)
were in the area $-10^{\circ}<l<40^{\circ}$ and $|b|<3^{\circ}$. Of these,
216 are SiO detections, giving a detection rate of 63 \%. Discussions of
individual objects and the spectra of SiO detections are given in the
Appendix.



\begin{figure}
  \begin{center}
    \FigureFile(90mm,90mm){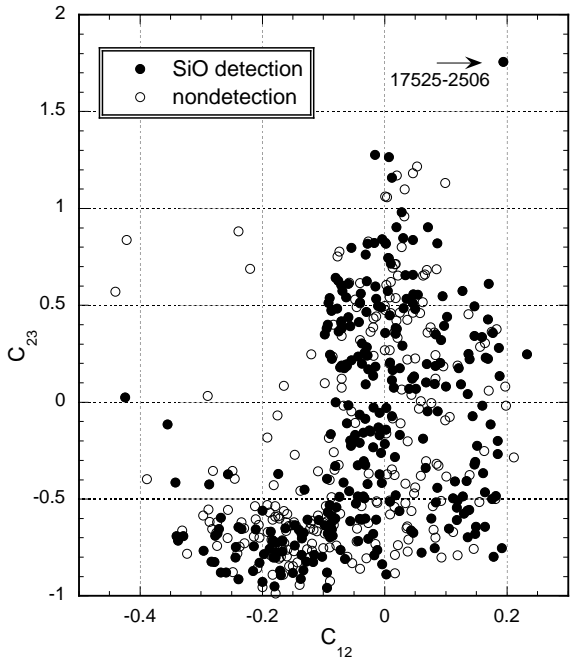}
  \end{center}

\caption{IRAS two-color diagram of the observed sources. Filled and unfilled
circles indicate the SiO detections and nondetections. 
IRAS 17525-2506 shows extreme colors as
an AGB star (see text). Two other objects,
16342$-$3814 and 18596+0315, have extreme colors and are outside this
figure.}

\label{fig:two-color.diagram}

\end{figure}
\begin{figure}
  \begin{center}
    \FigureFile(75mm,65mm){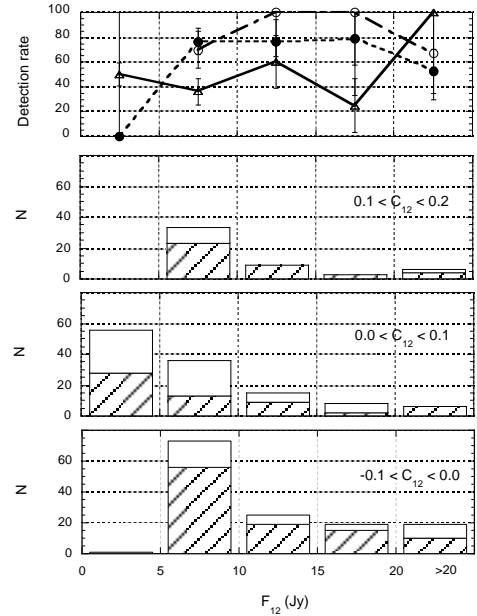}
  \end{center}

\caption{Detection probability (top line graph) and histograms of flux
density (lower three).  The broken,  solid, and  dash-dot  lines in the top
line graph indicate the subsamples of objects with $-0.1<C_{12}<0.0$,
$0.0<C_{12}<0.1$, and $0.1<C_{12}<0.2$, respectively. The shaded and unshaded
histograms indicate the SiO detection and nondetection, respectively.  The
detection rate for objects with $0.0<C_{12}<0.1$ appears to be lower than
that for objects with other colors, but this is an observational selection
effect (see main text).

}\label{fig:histogram}
\end{figure}

Figure 2 shows the two-color diagram of the observed sources.  In this
figure, we see a sudden decrease of objects at $C_{12}<-0.1$. However, this
is totally an observational selection effect, because the color-selection
criteria were slightly different for the high galactic latitude stars
($|b|>3^{\circ}$) (because contamination to the IRAS 60 $\mu$m flux density
by molecular clouds is negligible at the high latitudes; see \citet{izu99}),
In the Galactic disk area, $l<40^{\circ}$ and $|b|<3^{\circ}$, contamination
of the IRAS 60 $\mu$m flux density by dust clouds causes an increase of
SiO detectable objects at $C_{23}>0$, which is a region not populated by AGB
stars (\cite{van88}).  This is well known and discussed in various
papers (\cite{izu99}).

Figure 3 shows a comparison of SiO detection rates in the subsamples of the
color ranges, $-0.1<C_{12}<0.0$, $0.0<C_{12}<0.1$, and $0.1<C_{12}<0.2$.
Apparently, the detection rate for objects in the color range
$0.0<C_{12}<0.1$  is lower than the detection
rates for objects in the other color ranges. However, this is due to a
selection bias in the sources that were observed (The objects in IRAS color
range $0.0<C_{12}<0.1$ were selected and observed before 2000, when the MSX
catalog was not available, while those in the IRAS color ranges,
$-0.1<C_{12}<0.0$ and $0.1<C_{12}<0.2$ were selected after 2000, making use
of the MSX catalog). The positional accuracies of the MSX sources ($\sim
3''$) were much improved over those of the IRAS Point Source
catalog ($\sim 10''$).  Furthermore, the known young stellar objects are
excluded (using the SIMBAD database, MSX images, and 2MASS images) in the
observations after 2000. Thus, because of the improvements in the source
selection and positions, we have SiO detection rates as high as $\sim 80$ \%
for the objects in the range $-0.1<C_{12}<0.0$ and $0.1<C_{12}<0.2$.

\section{Discussion}

Note that the present SiO survey was made with pointed observations and is
not an unbiased survey as is the case with the OH 1612 MHz survey
(\cite{sev01}).  A possible bias in the detection rate may be caused by the
source selection process using two different catalogs (IRAS and MSX), as
noted in the previous section. Nevertheless, the present SiO maser sample is
large enough as a basic database of radial velocities of evolved stars in
the inner Galactic disk. With the above caution in mind, we discuss the
radial velocity data.

\subsection{l-v diagram}

The left panel of figure 4 shows  a longitude--velocity diagram for the SiO
detected sources in this paper (open circles). For the purpose of comparing
this figure with the OH $l$--$v$ diagram (right panel), we added 277 SiO
detected sources (squares) within the color range, $0.0<C_{12}<0.1$, in
$-10^{\circ}<l<25^{\circ}$ and $|b|<3^{\circ}$ from previous papers
(\cite{izu99}; ; \cite{deg00a}; \cite{deg00b}). The addition makes the
total number of SiO detections 493, similar to the total number of OH 1612
MHz detections (583 in the same region).
The $l$--$v$ diagrams are
overlaid on the CO $J=1$--0 map (\cite{dam01}), which gives an approximate
idea of the range of velocities due to the Galactic rotation.

The overall patterns in the SiO and OH $l$--$v$ diagrams resemble each
other: the SiO $l$--$v$ map (figure 4) exhibits two areas with no detections
which are indicated as ellipses H1 and H2.  The same empty regions can also
be seen in the OH $l$--$v$ diagram (the right panel of figure 4) [also in
the 86GHz SiO $l$--$v$ diagram of \citet{mes02}]. The presence of such vacant regions 
in the  $l$--$v$ diagram seems to  indicate that a spatial density of IRAS sources 
is considerably low in inter-arm regions (for H1), and also  irregular even in spiral arms (for H2), 
though the IRAS source density, when integrated on the line of sight,  does not vary much
along longitude. 

Furthermore, we can recognize subtle differences between the two panels; the
SiO sources seem to avoid the 3 kpc arm feature, which is shown as the
broken line in the left panel of figure 4, but this void does not seem to
appear strongly in the OH map (right panel). In addition, the OH
distribution seems to correlate strongly with the molecular ring feature
which extends approximately between ($l$,$V_{\rm lsr}$)=(10$^{\circ}$, 20 km
s$^{-1}$) and (25$^{\circ}$, 60 km s$^{-1}$). However, the SiO map does not
seem to show such a very strong concentration towards the molecular ring. 
Rather, it shows a high concentration of sources near the upper edge of the
molecular ring feature at ($l$,$V_{\rm lsr}$) $\sim $(19$^{\circ}$, 70 km
s$^{-1}$). This impression, i.e., that there is less concentration of SiO
sources towards the molecular ring, comes mainly from the sparse
distribution of SiO detections around ($l$,$V_{\rm lsr}$) $\sim
$($17\pm1.25^{\circ}$, 35$\pm 25$ km s$^{-1}$), whereas the OH $l$--$v$
diagram exhibits a high density of detections at the same region. However,
it turns out that the impression is not correct; we count, in fact, the SiO
sources, 18204$-$1344 (=IRC $-$10414), and the other three (18162$-$1422,
18182$-$1447, and 18187$-$1430, detected by SiO in \citet{mes02}) falling in
this spatial and velocity range. These are simply dropped from the present
survey list due to color-selection criteria, and not shown in figure 4.



The filled triangles in the OH $l$--$v$ diagram (the right panel in figure
4) indicates OH 1612 MHz double-peak sources without NIR counterparts within
5$''$ of their positions.  More than half of these objects are considered to
be gas clouds in star forming regions (not evolved stars), which often emit
dominant OH 1665 MHz masers (of course, not all of them necessarily follow
this rule ; \cite{cas99}).  We can see that these filled triangles (SFR
candidates) are concentrated strongly along the molecular ring feature in
the right panel of figure 4 (and with the nuclear disk, which is in the
range $|l| \lesssim1^{\circ}$ and $|V_{\rm lsr}| \lesssim 200$ km s$^{-1}$).


The sample of OH 1612 MHz sources with doubly peaked spectra involves a
variety of objects: AGB and post-AGB stars, planetary nebulae as well as gas
clouds in star forming regions (\cite{cas99}).  On the other hand, the SiO
sample involves purely AGB stars (very few post-AGB and young stellar
objects). It is well known that the gas dynamics differs considerably
from the stellar dynamics in the Galaxy (e.g., \cite{bin91}).  Counting the
OH 1612 MHz objects without NIR counterparts in the 2MASS database,
we estimate that more than 10 \% of the \citet{sev97}'s OH 1612 MHz objects
are likely to be gas clouds in SFRs (of course, the percentage is somewhat
uncertain, depending on the color selection criteria and the limiting
magnitudes in the NIR bands).

Therefore, we concluded that the OH and SiO surveys are looking at slightly
different (stellar) types of stars. Especially, exclusion of SFR OH 1612 MHz
sources from the OH sample is crucial when making comparisons between the OH
and SiO samples.



\begin{figure*}

\vspace{-0.6cm}

\begin{center}
\FigureFile(160mm,170mm){fig4.eps}
\end{center} 
\vspace{-4.2 cm} 
\caption{ 
SiO (left) and OH (right) longitude--velocity diagram.  In the left
panel, open and crossed circles indicate the sources below and above
$|b|=10^{\circ}$, which were detected in the present paper. The open squares
are the SiO sources detected by the previous SiO surveys
(\cite{izu99};\cite{deg00a}; \cite{deg00b}). The OH data (right panel)
were taken from \citet{sev97} and \citet{sev01}, and open circles and filled
triangles indicate the objects with and without NIR counterparts. The places
where SiO masers are absent, H1 and H2, are indicated by thick ellipses. 
The broken lines indicate the 3 kpc arm.  \vspace{-0.3 cm}
}

\end{figure*}

\subsection{2MASS identifications of IRAS sources and estimation of
distances}

Identifications of IRAS sources with 2MASS  near-infrared objects were made
in terms of $H-K$ colors and $K$ magnitudes (see detail in
\cite{deg98}).  Table 4 gives a summary of the identifications of the IRAS
sources with SiO detections, listing IRAS name, MSX 5CG name, right
ascension and declination in J2000, $J$, $H$, and $K$ magnitudes, blending flag, and the
distance from the Sun. Near-infrared identifications of the IRAS sources
were made successfully for all the objects except one (IRAS 17525$-$2506).
The distances were calculated from total infrared fluxes, which are obtained
from $J$ (1.25 $\mu$m), $H$(1.65 $\mu$m), $K_S$(2.16 $\mu$m) magnitudes and IRAS
12, 25, and 60 $\mu$m flux densities, assuming a constant stellar luminosity
of $L=8 \times 10^3 \; L_{\odot}$.  A standard model for the dust extinction
distribution (\cite{una98}) was used to get the extinction correction in the
galactic disk [see \citet{deg98} in detail].  Plots of $J-K$ and $K$ against
$H-K$ of the identified objects are shown in figure 5, where the broken
lines in the left and right panels indicate the expected $J-H$ and
$K$ magnitude variations due to interstellar and circumstellar extinction,
respectively. The squares in the lower-left indicate the position of an M6
III star without extinction (at a distance of 8 kpc in the right panel).  In
the left panel, most of the points (except stars with low photometric
quality which are indicated with open triangles) fall near the broken line
as expected. In the right panel, most of points fell below the broken line,
suggesting that the distances of these stars are smaller than 8 kpc.


\begin{figure*}
\vspace{-1cm} 
\begin{center} 
\FigureFile(160mm,170mm){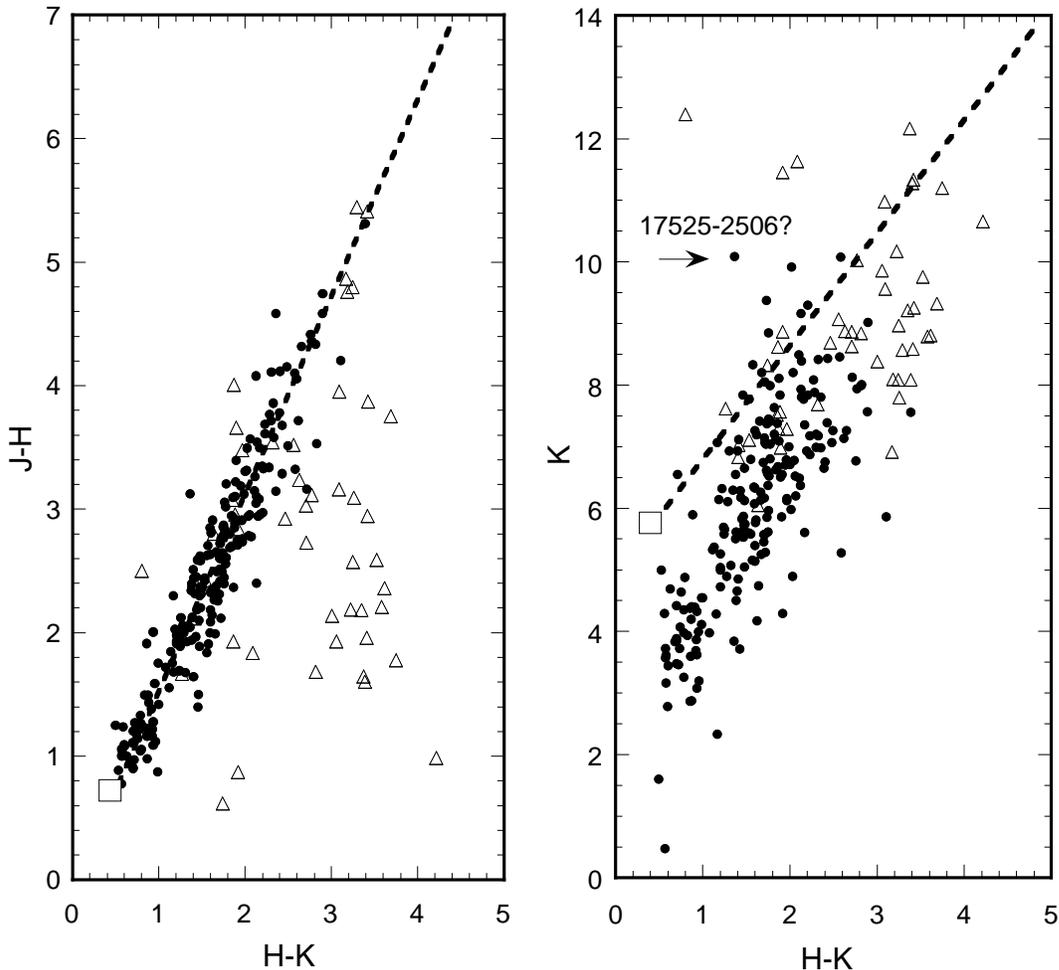}
\end{center} 
\vspace{-1cm} 
\caption{Plots of  $J-H$ (left panel) and $K$ (right panel) against $H-K$
for the 2MASS identified stars. Filled circles and open triangles indicate
the objects with high and low photometric qualities, respectively.  The
squares in the lower-left indicates the position of a M6 III star without
extinction (at a distance of 8 kpc in the right panel).  The identification
of IRAS 17525$-$2506 is not certain (see Appendix). \vspace{-0.1cm} }

\label{fig:l-v.diagram} 

\end{figure*}

Using the distances that we obtained, we mapped the positions of the SiO
detections on the galactic plane (in the first quadrant), as shown in figure
6.  Because the present SiO data are rather limited in their longitude and
color ranges, we added previously known SiO detected objects (squares)
within $l<70^{\circ}$ from previous papers (\cite{izu99}, \cite{nak03},
etc.).  A spiral model (figure 1 of \cite{tay93}), which indicates spiral
arms and HII regions, was overlaid on the figure. We can recognize large
concentrations of sources, one in the area around (3, 0.5) and the other
near the line connecting the points (5, 3) and (0, 8). The latter
concentration of sources corresponds to the Bulge bar, and has been
discussed previously (for example, \cite{dwe95}; \cite{nik97}); this feature
was already recognizable in the smaller sample of narrow $C_{12}$ color
range objects (see figure 6 of \cite{deg02}). The other concentration of
sources, around (3, 0.5), is located near a dense area of HII regions,
corresponding to the ``molecular ring feature" in the $l$--$v$ diagram
(\cite{dam01}). Note that no large concentration was found towards the ``3
kpc ring" feature except for the starting point of the feature around (4.8,
1), which corresponds to the tip of the Bulge bar.


\begin{figure*}
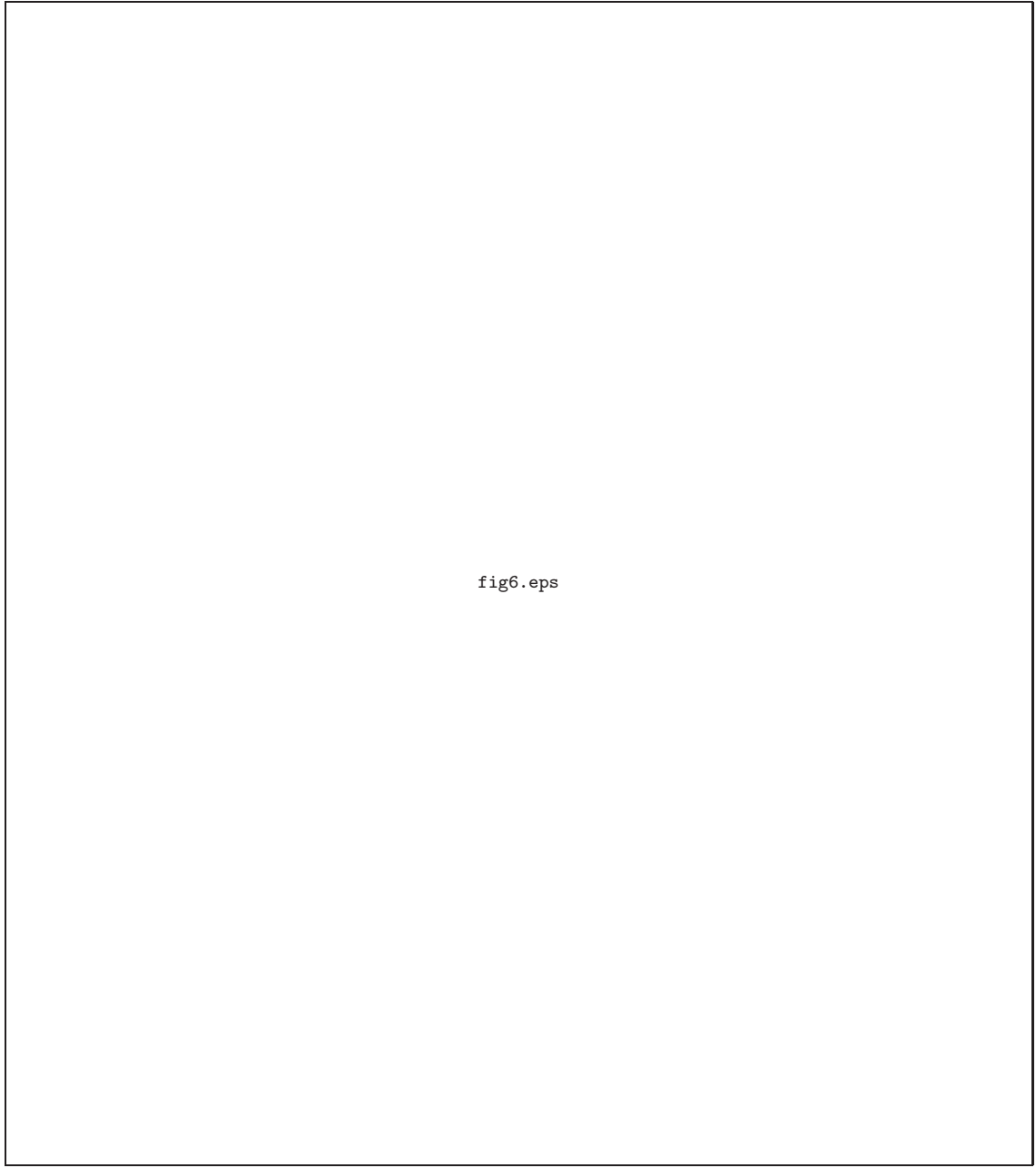

\vspace{-0.3cm} 
\begin{center} 
\FigureFile(150mm,170mm){fig6.eps}

\caption{Face-on view of the first quadrant of the Galaxy. The positions of
the detected SiO sources are shown by open circles (from this paper) and
squares (from the other papers quoted). The objects at $|b|>10 ^{\circ}$ are
indicated as circles with crosses. The sun is at the origin (shown as
$\odot$) and the galactic center is at (8,0). A spiral model of the Galaxy
(\cite{tay93}) is overlaid (thin curves for spiral and thin filled circles
for HII regions).
} 

\end{center} 
\end{figure*}

\subsection{Pattern speed of SiO maser sources}

From the SiO radial-velocity data, we can calculate a pattern speed of the
Galaxy without, in principle, the use of dynamical models (\cite{tre84}). 
However, because we are seeing the Galaxy in the edge-on direction, the
method of computation must be modified from the original (\cite{deb02}).
Furthermore, because the present sample is not evenly distributed in
longitude with respect to the galactic center, we used a direct method,
which is described as follows:

We only consider projected quantities onto the galactic plane. 
With the pattern speed, $\Omega _{\rm P}$, the surface density 
can be written as a function of time in galacto-centric coordinates
as $\sigma(t, r, \phi) = \sigma(0, r,  \phi - \Omega _{\rm P}\, t) $.
The derivative with respect to time, $\partial \sigma / \partial  t$, can be 
replaced by
$ - \Omega _{\rm P} \partial \sigma /\partial \phi $
in the continuity equation, leading to a pattern speed
\begin{equation}
\Omega _{\rm P} = \overline{ div (\sigma  {\bf v}) }  \;  / \;
\overline{\partial \sigma /\partial \phi  } ,
\end{equation}
where the bar indicates the surface integral over the disk area
concerned. By Gauss's theorem, the surface integral can be changed to the
integral along the curve enclosing the area. To make use of the present
observed data, which are discrete quantities in density and velocity, they
have to be made continuous.  For this purpose, we approximate the discrete
density/velocity profile by a sum of Gaussian-shape profiles as follows
\begin{equation}
\sigma (x,y) = \sum_{i} \; exp\{ [(x-x_i)^2+(y-y_i)^2]/a^2 \}/(\pi a^2),
\end{equation}
and the velocity fields by 
\begin{equation}
{\bf v}(x,y) = \sum_{i} \; {\bf v}_i \;  exp\{ [(x-x_i)^2+(y-y_i)^2]/a^2 \}/(\pi a^2),
\end{equation}
where the summation must be taken over all the sampled stars with known
radial velocities. In this process, a smoothing parameter, $a$, is required
to bridge vacant parts of the density profile in the concerned area (keeping
as close as the true density as possible).  We show later that the pattern
speed that we obtain does not depend strongly on this parameter when 0.2 kpc
$< a < 0.5$ kpc.

In the first quadrant of the Galactic plane, we took four fan-shape areas
between radii 1 and 2.5 kpc, 2.5 and 4 kpc, 4 and 5.5 kpc, and 5.5 and
7 kpc for integration, as illustrated in figure 7. The inflowing and
outflowing mass fluxes at the boundary must be computed. For computing the
inflow rate at $l=0^{\circ}$, we assume that all the particles at
$l=0^{\circ}$ move with a tangential velocity of $v_t= 220$ km s$^{-1}$
(because tangential velocities of the particles are unknown), which
corresponds to the galactic rotation (with a flat rotation curve).  This
assumption is satisfied only in an approximate sense.  Moreover, we assume
that the mass inflow rates at the $l=70^{\circ}$ boundary are negligibly
small. We used 547 (216 from this paper) SiO detected objects with color
$-0.1<C_{12}<0.2$ in the strip
$-10^{\circ}<l<70^{\circ}$ and  $|b|<3^{\circ}$.
The luminosity distances were calculated by the same method for all the
sources.  Because the present sample involves objects with distances mostly
below 7 kpc, we integrate only over the surface areas between 1 and 7
kpc.

Next, we calculated the inflow and outflow rates on the inner and outer arcs
(for example, on arc BI and JA, in figure 7) of the fan-shape area ABIJ from
the observed radial velocities (from equations 2 and 3).  Adding the inflow
rate from the $l=0$ boundary (for example, the line AB), and dividing them
by the average density differences at the boundaries (the denominator in
equation 1), we obtain the pattern speeds in the four fan-shape areas
between 1 and 7 kpc. The results are shown in figure 8.  The lower panel
shows the pattern speed in each fan-shape area between $D=1$ and 7 kpc. The
pattern speed obtained between 1 and 5.5 kpc for $a=0.4$ kpc (the area,
BEFI, adding the 3 separated fans into one) is

\begin{equation}
\Omega _{\rm P} = 21 \;  (\pm 13)  \; {\rm km \; s}^{-1} \;  {\rm kpc}^{-1}, 
\end{equation}
where the value between the parentheses is the root mean square deviation
from the average in the three areas. We calculated the pattern speed for
$a=0.05$ kpc to $a=0.5$ kpc in 0.05 kpc steps. The computed pattern
speed varies from 19 ($a=0.25$ kpc) to 32 ($a=0.05$ kpc) km s$^{-1}$
kpc$^{-1}$. Here, we take a positive pattern speed to be
prograde to the galactic rotation. The variation of the surface density,
$\sigma$, at $l=0^{\circ}$ as a function of radius is shown in the upper
panel of figure 8. When $a>0.2$ kpc, the surface density seems adequately
smoothed. When the smoothing parameter $a$ is varied between 0.2 and 0.5 kpc,
$\Omega _{\rm P} $ only changes between 19 and 24 km s$^{-1}$ kpc$^{-1}$.
Thus we consider that the value given in equation 4 is reasonable
for the pattern speed of the SiO maser sources in the distance range
$D=1$--5.5 kpc in the present sample.

As shown in figure 7, the pattern speed tends to increase with decreasing
radius from the Galactic center, though the number of SiO sources sampled at
large distances (beyond 6 kpc) is insufficient. The pattern speed of the
area, ABIJ ($D=$5.5--7 kpc), approaches 60 km s$^{-1}$ kpc$^{-1}$, (45
kms$^{-1}$ kpc$^{-1}$ for $a=0.25$ kpc, and 64 km s$^{-1}$ kpc$^{-1}$ for
$a=0.5$ kpc), This behavior of increasing pattern speed toward the Galactic
center does not appear to be understandable in terms of a single pattern
speed model for the Galaxy.

\begin{figure}
\begin{center}
\FigureFile(100mm,100mm){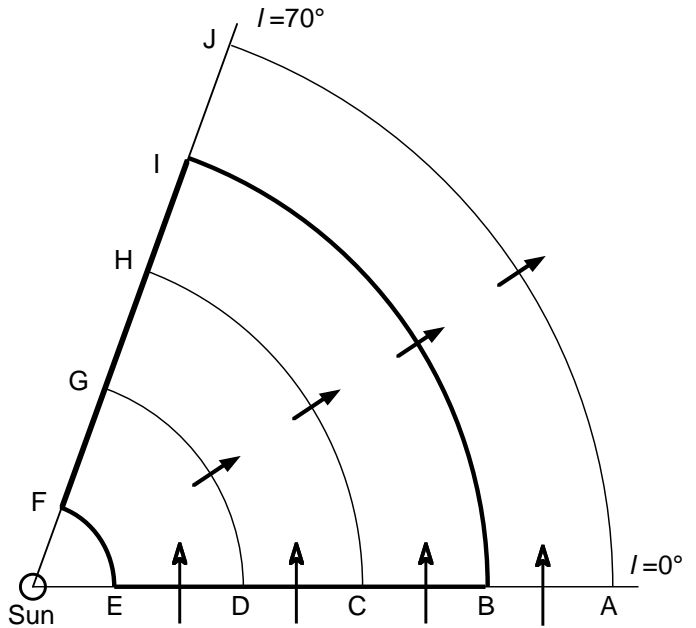}

\caption{Areas for integration. This figure indicate the first quadrant of
the Galaxy, approximately the same area as in figure 6. The arrow indicates
the flow of stars. The areas enclosed by ABIJ ($D=5.5$--7 kpc), BCHI
($D=4$--5.5 kpc), CDGH ($D=2.5$--4 kpc), and DEFG ($D=1$--2.5 kpc), are used
for numerical integration. The area enclosed by BEFI shown in thick outline is
used for taking the average.
} 

\end{center}
\end{figure}

Previously obtained values of the pattern speed of the Galaxy are widely
different: for example, 19 km s$^{-1}$ kpc$^{-1}$ from gas-dynamical
modeling of CO and HI near the galactic bar (\cite{wad94}), 53 km s$^{-1}$
kpc$^{-1}$ from the analysis of Hipparcos late-type stellar data in the
solar neighborhood (\cite{deh99}), 59 km s$^{-1}$ kpc$^{-1}$ from the
OH/IR sources (\cite{deb02}), 20  km s$^{-1}$ kpc$^{-1}$ from dynamical modeling 
of the K-band spiral pattern
(\cite{mar04}). The value of the pattern speed obtained in the
present paper, $\sim 20$ km s$^{-1}$ kpc$^{-1}$ gives the radius of
co-rotation (to the galactic rotation) to be approximately 11 kpc, which is
unreasonably large. However, a more recent hydrodynamic simulation with two
pattern speeds (\cite{bis03}) gives a good fit of the calculated gas pattern
to the CO $l$--$v$ diagram, resulting in
$\sim$60 km s$^{-1}$ kpc$^{-1}$ and $\sim$20 km s$^{-1}$ kpc$^{-1}$ for the
bar and spiral-arm pattern speeds, respectively. Thus, the tendency of the
pattern speed to increase towards the Galactic center, which is found in
this paper, can be interpreted as a manifestation of the two (inner fast and
outer slow) pattern speeds for the Galactic bar and spiral arms.  In this
case, the corotation radius of the bar appears around 3.5 kpc and the radius
of the inner Lindblad resonance of the two-armed spiral pattern around 3 kpc
(\cite{bin87}; and see fig. 3 of \cite{bis03}); this coincidence can
possibly create the 3 kpc arm as described in \citet{bis03}.

\begin{figure}
  \begin{center}
  \FigureFile(90mm,90mm){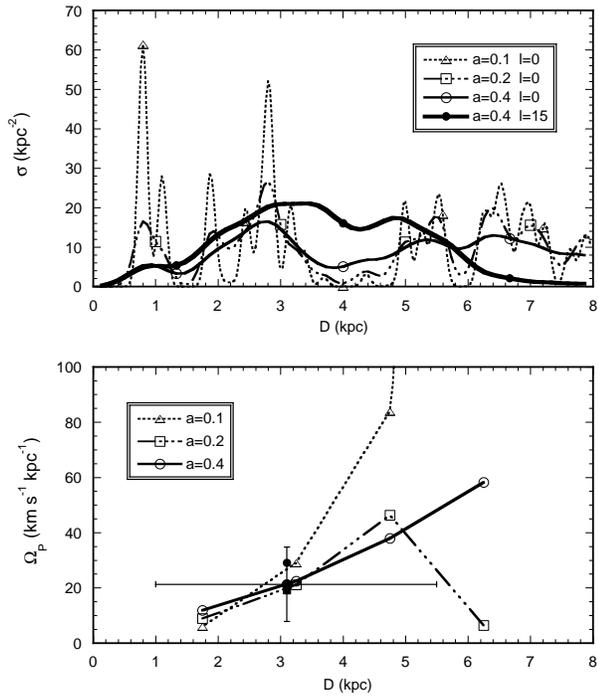}

\caption{Dependence of the results on the smoothing parameter, $a$. The
upper panel shows the density profile along $l=0 ^{\circ}$, and the lower
panel shows the pattern speed averaged over each 1.5 kpc in distance.
The vertical error bar in the lower panel indicates the rms deviation of the
three pattern speeds in the different areas, but the horizontal ``error bar"
indicates the inner and outer radii for the area used for the integration.
} 

\end{center}
\end{figure}

Though the pattern speed obtained in this paper does not depend on any
dynamical models, the computation still involves several uncertain
quantities such as the distances to the sources (we use luminosity
distances), a mass inflow rate at $l=0^{\circ}$ given by a rotation curve
($v_0=220$ km s$^{-1}$), and a value of the smoothing parameter (we took
$a=0.4$ kpc as a representative value).  Fortunately, mild variation of
these parameters does not change the pattern speeds by large factors.
Thus, we believe that the pattern speeds presented in this paper are
close to the true ones.

\section{Conclusion}

We have surveyed 403 IRAS sources in the galactic disk and obtained 254 (239
new) detections in the SiO $J=1$--0 $v=1$ and 2 lines. The detections of SiO
masers in IRAS sources with unknown characteristics constrain them to be
O-rich evolved stars. The SiO $l$--$v$ diagram in the area
$-10^{\circ}<l<40^{\circ}$ shows a lesser degree of concentration of SiO
sources towards the molecular ring feature than does the OH 1612 MHz source
$l$--$v$ map, suggesting some difference of stellar type between SiO and OH
objects.

We also identified all the SiO detected objects with 2MASS stars.  We mapped
the objects on the Galactic plane using luminosity distances based on the
near- and mid-infrared flux densities. Combined with the radial-velocity
data, we found the pattern speed of the SiO maser sources to be $\Omega
_{\rm P}=21 \; (\pm 13)$ km s$^{-1}$ kpc$^{-1}$ between galactocentric
distances of 1 and 5.5 kpc, and $\Omega _{\rm P}\sim 60$ km s$^{-1}$
kpc$^{-1}$ between distances 5.5 and 7 kpc.

We are grateful to graduate/undergraduate students, Mrs. Y. Eto, F. Ieda, M.
Ioroi, R. Tamura, K. Watarai, and Ms. H. Fukushi for assistance with the
observations during long runs. This research makes use of the SIMBAD
database operated at CDS, Strasbourg, France, as well as data products from
the Two Micron All Sky Survey, which is a joint project of the University of
Massachusetts and the Infrared Processing and Analysis Center/California
Institute of Technology, funded by the National Aeronautics and Space
Administration and the National Science Foundation. This research was partly
supported by Scientific Research Grant (C2) 12640243 of Japan Society for
Promotion of Sciences.

\section*{Appendix. Individual Objects}
We give SiO maser spectra for the detected objects
in figures 9a--9p, and discuss individual objects here.
\begin{itemize}

\item 15060+0947 (=FV Boo) :  This high-latitude star ($b=53.27^{\circ}$)
exhibits an interesting double-peak emission profile in SiO lines in May
2003 (see top left in figure 5b) at $V_{\rm lsr}=-8.3$ and 2.1 km $^{-1}$;
double peak SiO maser sources are relatively rare. \citet{lew02} listed this
object as a dead OH/IR star, in which the OH 1612 MHz maser emission
declined.  The previous OH 1612 MHz line detection of this star
(\cite{ede88}) was at $V_{\rm lsr}=-16.3$ and 1.8 km s$^{-1}$, resulting in
a stellar velocity of $V_{\rm lsr}=-7.1$ km s$^{-1}$. \citet{eng96} detected
the triply peaked H$_2$O maser emission at $V_{\rm lsr}=-13.4$, $-1.6$, and
3.0 km s$^{-1}$, and \citet{ita01} detected SiO emission at slightly larger
radial velocities in 2000 April, indicating a strong time variation. These
facts suggest that this star is near the end of the AGB phase with varying
mass loss rate.

\item  16156$-$2837  (=V932 Sco) : This is a very bright star on the 2MASS
$K$-band images ($K=2.78$). The {\it General Catalog of Variable Stars}
(\cite{kho85}) classified it as a variable star of Orion-type.  The IRAS
LRS class of this object is 27, implying that it has a silicate emission
feature. A previous OH maser search was negative (\cite{tel91}), but SiO
masing was detected for the first time in this paper.

\item  16342$-$3814 (=OH 344.07+5.84) and 18596+0315 (OH 37.12$-$.85): These
are post-AGB stars with cool circumstellar envelopes ($C_{12}>0.5$;
\cite{sah99}). Though H$_2$O masers have been detected at velocities
outside the OH double peak velocity range (\cite{zuc87} ; \cite{tel88};
\cite{gom94}), SiO maser searches have been negative, which is consistent
with the post-AGB star interpretation.

\item  16473$-$2528 (=AF Sco) :
This star was previously classified as a nova or cataclysmic variable, but
has now been identified as a Mira-type variable (\cite{due95}). The IRAS LRS
class is 29, indicating strong silicate features. Detections of the OH 1612
MHz and main lines (\cite{tel91}; \cite{lew95}), and of SiO masers in the
present paper, are all consistent with the Mira interpretation of this star.

\item  17269$-$2625  (=V2311 Oph): 
The IRAS LRS spectral class for this object is 41 (indicating an 11 $\mu$m SiC
feature), but \citet{syl99} found the 10 $\mu$m silicate emission with UKIRT
(=IRC$-$30300).  \citet{gro94} found the 18 $\mu$m silicate feature is
noticeable in the IRAS LRS spectrum and gave the spectral type as $\geq$M8. 
The SiO detection in this paper secures that this is an O-rich AGB star. No
OH masers have ever been detected.

\item  17367$-$2319 (=V545 Oph):
The OH 1612 MHz search was negative (\cite{tel91}). This is a strong IRAS source ($F_{12}=55.8$ Jy)
with LRS class 26 (silicate feature). \citet{gro99} detected CO $J=2$--1 emission from this star
at $V_{\rm LSR}=-18.7$ km s$^{-1}$, which coincides with the SiO velocity 
at $V_{\rm lsr}=-19.2$ km s$^{-1}$ in the present paper.

\item  17525$-$2506 (=OH 4.417 +.044) :
 This object occupies a unique position in the two-color diagram (figure 2),
because the IRAS Point Source Catalog gives a very high 60 $\mu$m flux for
it ($F_{60}$=1007 Jy, but $F_{12}$=11.3 Jy). However, this is due to
contamination by a surrounding young stellar object (IRAS 17522$-$2504 with
$F_{60}$=1007 Jy), in a star forming region. The positions of the OH 1612 MHz
maser (\cite{bec92}; \cite{sev97}) coincide with those of
MSX5CG004.4181+00.0438 within 4$''$.  A close look at 2MASS images toward
this object and at the DENIS database give several faint red stars around
the OH positions, which are equally dubious as candidates.  SiO masing from
this object is relatively strong (see figure 9d).

\item  17559$-$2848 (=V4336 Sgr) :
This is a relatively strong IRAS source ($F_{12}=29.3$ Jy) 
with LRS class 25 (silicate emission). This is  located in the Sgr I optical window
and  \citet{gla95} measured the period of NIR light variation of 557 d.
OH 1612 MHz double peaks were detected (\cite{tel91}; \cite{sev97})
resulting in a stellar velocity $V_{\rm lsr}=-39.1$ km s$^{-1}$, which is consistent with 
SiO detection at $V_{\rm lsr}=-40.0$ km s$^{-1}$ (present paper).

\item  17599$-$2603:
This object (MSX5C\_G004.4403-01.8761) is located close to the globular cluster, Terzan 10
($\sim 1.4'$). Considering the distance of this glubular cluster  (5.7 kpc; \cite{har96}),
we cannot deny a possibility of this relatively strong middle-IR source ($F_{12}=8.6$ Jy) 
to be a member of this glubular cluster. Because the radial velocity of this glubular cluster has not been
measured unfortunately, the SiO radial velocity ($V_{\rm lrs}=81.7$ km s$^{-1}$) cannot be used 
to judge the membership.

\item  18038$-$1614 :
The IRAS LRS class of this source is 14, indicating a featureless spectrum.
However, we can see silicate emission features at 9.8 and 18 micron in the raw 
LRS spectrum (\cite{oln86}). 
The SIMBAD database indicates that this is a carbon star,
but the detection of SiO masers (present paper) excludes such a possibility.
Searches for the CO $J=2$--1 and 1--0 lines and OH 1612 MHz emission 
were negative (\cite{gro02}; \cite{tel91}; \cite{les92}). 

\item 18210$-$1359  :
The 2MASS images give two equally red candidate stars: one 16$''$ east and
the other 20$''$ west of the IRAS position. The MSX map also gives two
point-like objects at similar positions. We observed and detected SiO masers
in the MSX point source 16$''$ east, MSX5CG017.3984$-$00.3836, which the
brighter of the two in $K$ magnitude ($K=$6.53 and $H-K=$2.06). The SIMBAD
assignment of this object (IRAS 18210$-$1359) to GAL 017.4$-$00.4 (supernova
remnant candidate) is inappropriate.

\item  18213$-$1739 :
2MASS images show a very red star [$R.A.=18^{\rm h}24^{\rm m}15.12^{\rm s}$,
$Dec=-17^{\circ}37'44.3''$, J2000, with $K=$11.20 and $H-K=$3.75]  within
3$''$ the IRAS position, while the DENIS database gives 2 brighter stars
with $K=8.70$ and 8.72 at almost the same position. The difference in the
$K$ magnitudes between 2MASS and DENNIS can be attributed to a time
variation.  The SIMBAD assignment of this IRAS source as an A2 star, HD
169195 (more than 1' away from the IRAS position), seems inappropriate.  SiO
masers were detected for the first time in this paper and the previous OH
search (\cite{tel91}) was negative.

\item  18417$-$0205 : \citet{fel02} classified this object
(ISOGAL$-$PJ184418.5$-$020153) as a YSO candidate with confidence level 0. 
SiO masers were detected during this program. The DENIS $K$ magnitude of the
identified star is 8.68, and the 2MASS database gives $K=7.84$ with
$H-K=1.89$, suggesting that it is a variable star. These facts are enough to
conclude that this is an evolved object.

\item 18588+0428 (=OH 38.101$-$0.125) :
This source has been classified as a post-AGB star in the SIMBAD database.
However, the detection of relatively strong SiO masers at $V_{\rm lsr}=52$
km s$^{-1}$, which is consistent with the center velocity of the OH maser
double peaks (\cite{lew94}; \cite{sev02}), excludes the possibility of a
post-AGB star. The IRAS LRS class is 39, indicating a silicate absorption
feature at 9.8 $\mu$m.  The 2MASS database gives a very red star ($K=8.57$
and $H-K=3.29$) near the IRAS position.

\item 19093$-$3256 (=V342 Sgr) :
This is a bright Mira ($P=372$ d)  with $F_{12}=318$ Jy and $K=1.3$ located out of
the Galactic plane at [$(l,b)=(4.79^{\circ}, -18.44^{\circ})$]. The SiO
$J=2$--1 $v=1$ (\cite{hai94}) and
$J=1$--0 $v=1$ and 2 masers (present paper)  have been detected
at $V_{\rm lsr}=40.8$ km s$^{-1}$, which is consistent with the  CO $J=1$--0
radial velocity at $V_c=37.7$ km s$^{-1}$ (\cite{nym92}). The OH maser
search was negative (\cite{tel91}).
\end{itemize}


\include{table1}
\include{table2c}
\include{table3}
\onecolumn
\setcounter{figure}{8}
\begin{figure}
\vspace{-1cm}
  \begin{center}
    \FigureFile(160mm,230mm){fig9a.eps}
  \end{center}
  \caption{a -- p. SiO $J=1$--0 $v=1$ and 2 spectra for the detected sources.} 
\end{figure}
\setcounter{figure}{8}
\begin{figure}
\vspace{-1cm}
  \begin{center}
    \FigureFile(160mm,230mm){fig9b.eps}
  \end{center}
  \caption{b.} 
\end{figure}
\setcounter{figure}{8}
\begin{figure}
\vspace{-1cm}
  \begin{center}
    \FigureFile(160mm,230mm){fig9c.eps}
  \end{center}
  \caption{c.} 
\end{figure}
\setcounter{figure}{8}
\begin{figure}
\vspace{-1cm}
  \begin{center}
    \FigureFile(160mm,230mm){fig9d.eps}
  \end{center}
  \caption{d.} 
\end{figure}
\setcounter{figure}{8}
\begin{figure}
\vspace{-1cm}
  \begin{center}
    \FigureFile(160mm,230mm){fig9e.eps}
  \end{center}
  \caption{e.} 
\end{figure}
\setcounter{figure}{8}
\begin{figure}
\vspace{-1cm}
  \begin{center}
    \FigureFile(160mm,230mm){fig9f.eps}
  \end{center}
  \caption{f.} 
\end{figure}
\setcounter{figure}{8}
\begin{figure}
\vspace{-1cm}
  \begin{center}
    \FigureFile(160mm,230mm){fig9g.eps}
  \end{center}
  \caption{g.} 
\end{figure}
\setcounter{figure}{8}
\begin{figure}
\vspace{-1cm}
  \begin{center}
    \FigureFile(160mm,230mm){fig9h.eps}
  \end{center}
  \caption{h.} 
\end{figure}
\setcounter{figure}{8}
\begin{figure}
\vspace{-1cm}
  \begin{center}
    \FigureFile(160mm,230mm){fig9i.eps}
  \end{center}
  \caption{i.} 
\end{figure}
\setcounter{figure}{8}
\begin{figure}
\vspace{-1cm}
  \begin{center}
    \FigureFile(160mm,230mm){fig9j.eps}
  \end{center}
  \caption{j.} 
\end{figure}
\setcounter{figure}{8}
\begin{figure}
\vspace{-1cm}
  \begin{center}
    \FigureFile(160mm,230mm){fig9k.eps}
  \end{center}
  \caption{k.} 
\end{figure}
\setcounter{figure}{8}
\begin{figure}
\vspace{-1cm}
  \begin{center}
    \FigureFile(160mm,230mm){fig9l.eps}
  \end{center}
  \caption{l.} 
\end{figure}
\setcounter{figure}{8}
\begin{figure}
\vspace{-1cm}
  \begin{center}
    \FigureFile(160mm,230mm){fig9m.eps}
  \end{center}
  \caption{m.} 
\end{figure}
\setcounter{figure}{8}
\begin{figure}
\vspace{-1cm}
  \begin{center}
    \FigureFile(160mm,230mm){fig9n.eps}
  \end{center}
  \caption{n.} 
\end{figure}
\setcounter{figure}{8}
\begin{figure}
\vspace{-5cm}
  \begin{center}
    \FigureFile(110mm,130mm){fig9o.eps}
  \end{center}
  \caption{o.} 
\end{figure}
\end{document}

%% file: table1.tex
\newcommand{\m}{$\dag$}
\small
\tabcolsep 1pt


%% file: table2c.tex
\tabcolsep 6pt
\begin{table}[t]
\caption{List of nagtive results}
\begin{tabular}{lccr}
\hline
      &  $v=1$ & $v=2$ &     \\
 IRAS name & rms & rms & Obs.date\\
      & (K) & (K) & ($yymmdd.d$)\\
\hline
 13466$-$3512 & 0.062 &0.069 &010410.0\\
 15191$-$3905 & 0.087 &0.092 &010410.0\\
 15589$-$2850 & 0.103 &0.088 &030526.0\\
 16039$-$2040 & 0.168 &0.135 &030521.0\\
 16275$-$2638 & 0.124 &0.091 &030523.0\\
\hline
 \end{tabular}
\end{table}

\setcounter{table}{1}
\tabcolsep 6pt
\begin{table}[t]
\caption{(Continued)}
\begin{tabular}{lccr}
\hline
      &  $v=1$ & $v=2$ &     \\
 IRAS name & rms & rms & Obs.date\\
      & (K) & (K) & ($yymmdd.d$)\\
\hline
 18290$-$1200 & 0.100 &0.096 &020523.2\\
 18294$-$0400 & 0.356 &0.442 &030516.2\\
 18303$-$0519 & 0.066 &0.062 &990525.1\\
 18307$-$0632 & 0.124 &0.114 &970124.6\\
 18310$-$0603 & 0.104 &0.138 &030516.2\\
\hline
 \end{tabular}
\end{table}
 
\setcounter{table}{1}
\tabcolsep 6pt
\begin{table}[t]
\caption{(Continued)}
\begin{tabular}{lccr}
\hline
      &  $v=1$ & $v=2$ &     \\
 IRAS name & rms & rms & Obs.date\\
      & (K) & (K) & ($yymmdd.d$)\\
\hline
 18425$-$0442 & 0.050 &0.045 &970522.9\\
 18425$-$0607 & 0.082 &0.070 &980513.2\\
 18430$-$0634 & 0.131 &0.122 &020315.3\\
 18440$-$0241 & 0.161 &0.135 &980409.3\\
 18440$-$0613 & 0.108 &0.090 &030526.1\\
\hline
\end{tabular}
\end{table}

\setcounter{table}{2}
%

%% file: table3.tex
\small
\tabcolsep 5pt
\begin{longtable}{lcrrrrrcr}
\caption{2MASS and MSX identification of IRAS sources with SiO masers}
\hline
IRAS name & MSX5C name & $RA$(J2000) & $Dec$(J2000) & $J$ & $H$ & $K$ & $Bflag$ & $D_L$ \\
          &         & ($\circ$)   & ($\circ$) & (mag) & (mag) & (mag) & & (kpc) \\
\hline		  
\endfirsthead		  
\hline		  
IRAS name & MSX5C name & $RA$(J2000) & $Dec$(J2000) & $J$ & $H$ & $K$ & $Bflag$ & $D_L$ \\
          &         & ($\circ$)   & ($\circ$) & (mag) & (mag) & (mag) & & (kpc) \\
\hline		  
\endhead
\hline
\endfoot
\hline
\multicolumn{8}{l}{$\dagger$ indicates that the identification is uncertain.} 
\endlastfoot
14251$-$3246  & ---                 & 217.0387 & $-$33.0010 & 8.680 & 7.282 & 5.830 &111& 2.50\\
14402$-$3645  & ---                 & 220.8292 & $-$36.9689 & 6.025 & 5.119 & 4.421 &111& 2.87\\
15060$+$0947  & ---                 & 227.1074 &     9.6051 & 5.450 & 4.520 & 3.836 &111& 1.12\\
15576$-$1212  & ---                 & 240.0990 & $-$12.3493 & 5.641 & 4.552 & 3.621 &111& 1.66\\
16038$-$2230  & ---                 & 241.7044 & $-$22.6438 & 4.978 & 3.742 & 3.161 &111& 1.66\\
16063$-$3227  & ---                 & 242.3742 & $-$32.5914 & 6.535 & 5.255 & 4.325 &111& 2.62\\
16091$-$1655  & ---                 & 242.9994 & $-$17.0539 & 5.977 & 5.103 & 4.116 &111& 2.15\\
16156$-$2837  & ---                 & 244.6974 & $-$28.7530 & 4.475 & 3.380 & 2.782 &111& 1.31\\
\end{longtable}